\documentclass{aa}  
\usepackage{graphicx,color}
\usepackage{txfonts}
\begin{document}
%
 \title{   Modeling  the X-ray light curves of  Cygnus X-3.    }
  \subtitle{    Possible role of the jet. }
  \author{O. Vilhu  
         \inst{1}
          \and
          D.C. Hannikainen\inst{2} 
  }

  \offprints{O. Vilhu}

   \institute{Dept of Physics, Division of Geophysics and Astronomy, University of Helsinki, P.O.Box 64, FI-00014 Helsinki, Finland\\
              \email{osmi.vilhu@helsinki.fi}
             \and
             Department of Physics and Space Sciences, Florida Institute of Technology, 150 W. University Blvd., Melbourne, FL 32901, USA \\
           \email{ddcarina@gmail.com}
                 }
   \date{Received  19.6.2012  ; accepted  16.12.2012 }
  \abstract
  {To address the physics behind the soft X-ray light curve asymmetries in Cygnus X-3, a well-known microquasar.} 
    {Observable effects of the jet close to the line-of-sight were investigated and interpreted within the frame of light curve physics. }
 {     The path of a hypothetical imprint of the jet,
 advected by the WR-wind,  was computed and  its crossing with  the line-of-sight during the binary orbit  determined. 
  We explore the possibility that physically this `imprint' is a formation
 of dense clumps triggered by jet bow shocks in the wind (``clumpy trail").  
 Models for X-ray continuum and emission line light curves were constructed using two absorbers:
mass columns along the line-of-sight of i) the WR wind and ii) the clumpy trail,  as seen from the compact star. These model light curves were compared with the observed ones from the $\it{RXTE}$/ASM (continuum) and $\it{Chandra}$/HETG (emission lines). }
{We show  that the shapes of the Cygnus X-3  light curves can be explained by the two absorbers using  the inclination and true anomaly angles of the jet   as derived in Dubus et al. (2010)  from  gamma-ray $\it{Fermi}$/LAT observations. The clumpy trail absorber is much larger for the lines than for the continuum. We suggest that the clumpy trail is a mixture of equilibrium and hot  (shock heated) clumps. }
{A possible way for studying jets in binary stars when the jet axis and the line-of-sight are close to each other is demonstrated. The X-ray continuum and emission line light curves of Cygnus X-3 can be explained by two absorbers: the WR companion wind  plus an absorber lying in the jet path (clumpy trail). We propose  that the clumpy trail absorber is due to dense clumps triggered by jet bow shocks. }
                                                                    
   \maketitle
%

\section{Introduction}

Cygnus X-3 (4U 2030+40, V1521 Cyg) is a high-mass X-ray binary (HMXB)
located at a distance of 9 kpc with a close binary orbit (P = 4.8 hour; Hanson et al. \cite{hanson};  Liu et al. \cite{HMXB}). 
The compact star is either a neutron star or a black hole and the companion is  most probably a WN5--7 type Wolf-Rayet (WR) star (helium star; van Kerkwijk et al. \cite{kerk92}, \cite{kerk96}). 
However, it is possible that the WR-phenomenon comes from the accretion disc wind, but to our knowledge no such detailed modeling  for this source exists.
 Cygnus X-3 is also a source of relativistic jets (e.g. Mioduszewski  \cite{mioduszewski}; Marti \cite{marti1}), thus associating Cygnus X-3 with the class of microquasars.

Massive winds are generally observed in Wolf-Rayet stars (Langer \cite{langer}; Crowther \cite{crowther}). 
This wind produces the well-observed orbital modulation of radiation from the compact star by asymmetric absorption during the orbit  along the line-of-sight.

The system inclination ($\it{i}$=the angle between the line-of-sight and the orbital axis) is unknown but probably small. Depending on inclination, the compact star mass is either large (3 -- 10 M$_{\sun}$  if $\it{i}$ = 30 deg) or smaller (1 -- 3 M$_{\sun}$ if  $\it{i}$ =  60 deg)  (Vilhu et al. \cite{vilhu}, see their Fig.~10).   
By modeling the X-ray spectra in all states, Hjalmarsdotter et al. (2009) concluded that the compact star is a massive black hole (30 M$_{\sun}$) indicating a very small inclination.
Koljonen et al. (2010) found six distinct X-ray states reminiscent of those seen in black hole transients, also supporting the black hole case.

Hot star winds are clumpy rather than homogeneous, consisting of dense condensations inside a rare gas (see papers in Hamann et al. \cite{hamann}).  
In Cygnus X-3  the clumps are photoionized by strong X-ray/UV radiation and are consequently highly radiative. 
Jets destroy clumps when colliding with them (Perucho \& Bosch-Ramon \cite{perucho2}). 
However, the opposite may also be true, such as in star formation regions where clumps instead form when a shock front collides with interstellar clouds (van Breugel et al. \cite{vanbreugel}).

     In  this paper, we explore the possibility 
that a bow shock from the jet indeed triggers the formation of clumps and/or makes the existing clumps denser and that these are advected by the WR-wind. 
The result is a ``clumpy trail'', a volume filled by the clumps and advected by the wind. 
In reality the formation of such a trail is far from being demonstrated but could in principle be investigated by radiative hydrodynamics (see Perucho and Bosch-Ramon \cite{perucho}).
The continuum and line absorptions are sensitive to density, 
and in different ways in a photoionised medium.  
Hence, one can expect observable effects  in X-rays if the line-of-sight from the compact star (X-ray source) crosses the clumpy trail region.  

We use the parameters derived by Dubus et al. (2009) for the jet inclination and true anomaly angles, 
      and compute the intersection of the clumpy trail with  the line-of-sight. 
Two different bow shock geometries are applied, and the continuum and emission line light curves modeled and compared with the observed ones. 


\section{Geometry}

 Dubus et al. (\cite{dubus}) modeled the $\it{Fermi}$/LAT gamma-ray orbital modulation of Cygnus X-3 by  inverse Compton scattering of WR photons on the jet using  two models for the compact star given in Szostek \& Zdziarski (2008): a neutron star with high orbital inclination (60 deg) and a black hole with small inclination (30 deg). 
Dubus et al. derived values for the jet inclination and true anomaly angles, and showed that the jet axis is close to the line-of-sight.
We used the black hole alternative, supported by the spectral studies of Hjalmarsdotter et al. (2009), Vilhu et al. (2009), and Koljonen et al. (2010). 
       The WR-wind is assumed  to be spherically symmetric with density depending on the distance r  from the WR-companion center as $r^{-\gamma}$
with $\gamma$ = 2.  

\begin{figure}
   \centering
 \includegraphics[width=9cm]{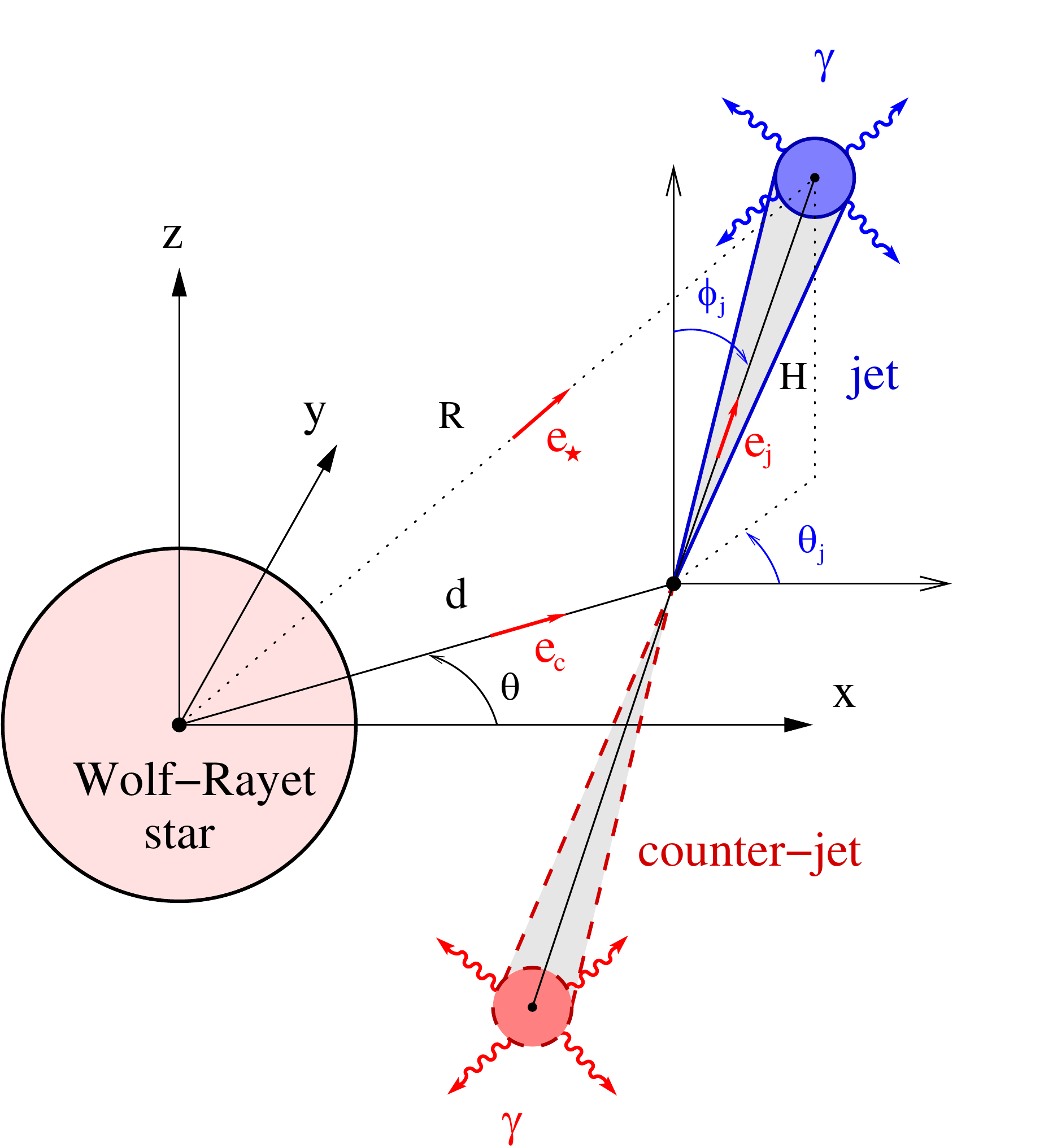}

      \caption{ The system geometry  (from Dubus et al. \cite{dubus}). The jet-angles used  are  
                     $\Theta$$_j$ = 319 deg and  $\phi$$_j$ = 39 deg. The line of sight  lies in a plane parallel to the yz-plane with  direction $\Theta$ (= $\Theta$$_{obs}$ ) = 270 deg and $\it{i}$ (=$\phi$$_{obs}$) = 30 deg.}
         \label{geometry}
   \end{figure}

The rectangular xyz-coordinate system was used where the WR star is in the center and the z-coordinate is parallel to the orbital axis ( see Fig.~\ref{geometry}).  
The  vector formalism and definitions  given in Dubus et al. (\cite{dubus}) were used.
The unit vectors of the jet axis and line-of-sight (observer)  in the xyz-coordinate system are:

\

e$_{jet}$ = (cos$\Theta$$_j$sin$\phi$$_j$, sin$\Theta$$_j$sin$\phi$$_j$, cos$\phi$$_j$) 

\

e$_{obs}$ = (0, -sin$\it{i}$, cos$\it{i}$).

\

Here $\Theta$$_j$  is the jet azimuth (true anomaly),  $\phi$$_j$  the jet polar angle (jet inclination=angle between the z-axis and the jet-axis) and $\it{i}$ the orbital inclination (angle between the z-axis and the  line-of-sight). 
The true anomaly is $\pm{90}$ deg at conjunctions (orbital phases 0 and 0.5).
For the  black hole case, Dubus et al. (2010) derived 39 deg  and 319 deg for   $\phi$$_j$  and    $\Theta$$_j$, respectively. Since the  observer has a permanent 270 deg true anomaly ($\Theta$$_{obs}$) and 30 deg inclination ($\phi$$_{obs}$),  the jet pointing is   close to the line-of-sight. 
Hence, Cygnus X-3 is more likely a ``microblazar" rather than a microquasar. 
Since the true jet anomaly is somewhat larger than that of the observer, the line-of-sight trails the jet during the orbit.

Besides the jet inclination  and azimuth, an  additional parameter in the modeling is the ratio H/d,  where d is the binary separation and H the height along the jet where high energy electrons are released and  Compton scattering takes place.  Dubus et al. (2010)  obtained a best value of 2.66 for this ratio.  
As an example,  for total masses 12 M$_{\sun}$ and  28 M$_{\sun}$  the binary separation d is 3.2 R$_{\sun}$  (2.2 $\times$10$^{11}$ cm) and 4.2  R$_{\sun}$  (4.2 $\times$10$^{11}$ cm), respectively  (the separation depends on  total mass as M$_{tot}$$^{1/3}$).

Numerical  models for microquasar jets  show a bow shock surrounding the jet  (Perucho \& Bosch-Ramon \cite{perucho}). 
We used  two simplified geometries  to represent the stationary shocked region surrounding the jet: i)  cone  shells with thickness  0.1d and  opening angles between 20--50 deg, and ii)    cylinders with thickness  0.1d and  radius  between 0.2d--0.3d and starting at distances between 0--1.0d from the compact star. 
Further, we assume that the  shocked region immediately transforms the wind material into 
clumps of equal density
with space density  N$_{wind}$/N$_{clump}$   advected radially from the WR-star by velocities 1000--3000 km/s.
N$_{wind}$ and N$_{clump}$ are the wind and clump densities, respectively. 
Hence, the wind mass is conserved in the process.
Note that this is just our assumption to explore the possibility of clumps without demonstrating that they really exist.
In numerical computations we used the  xyz-grid  divided into  70$\times$70$\times$70 pixels with pixel size 0.18d,
giving 12.6d for the box size.  

The cone geometry with opening angles less than 20 deg or larger than 50 deg did not produce any orbital modulation (Section 3), nor thinner tubes in the tube geometry. The shell thickness has a minimal impact since we are not interested in the absolute values of mass columns, but only relative values along the orbit.

\begin{figure}
   \centering
 \includegraphics[width=9cm]{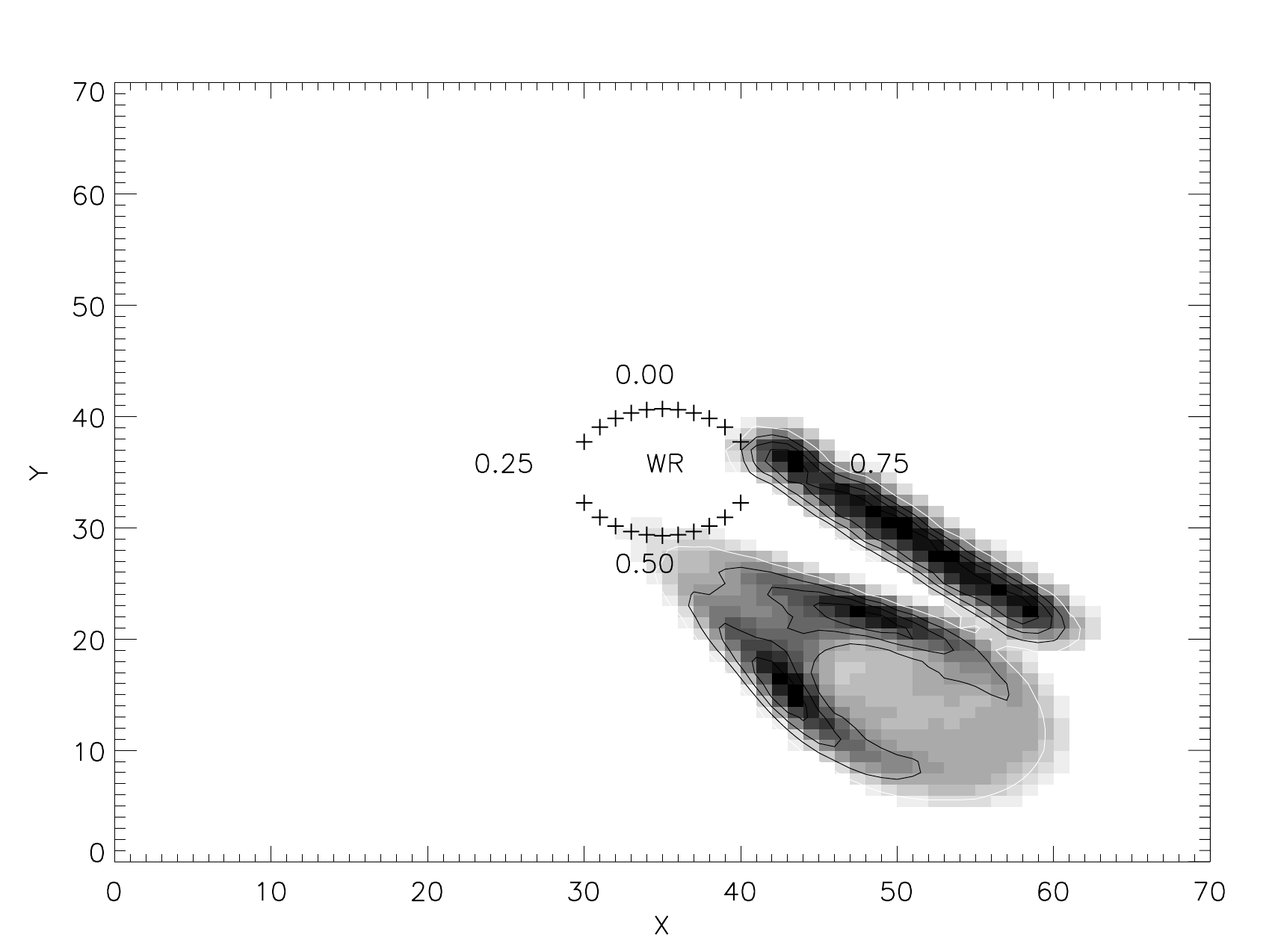}

      \caption{ Projection boundaries of bow shocks on the orbital xy-plane (pixel numbers marked)  
      for the two geometries used at orbital phases 0.35 (the cone geometry) and 0.85 (the tube geometry).  
      The WR-companion is in the center while the compact  star is moving in a  circular orbit around it (dotted circle). The observation direction is from below (see also Fig.~\ref{crossing}).}
         \label{jetimage}
   \end{figure}

The  geometries are shown  schematically in Fig.~\ref{jetimage}  as projections on the orbital plane.  
The WR-companion is in the center while the compact  star is moving in a  circular orbit around it (dotted circle wit four phases marked). 
The observation direction is from below (see also Fig.~\ref{crossing}). 
Note that the projections appear to end before reaching the grid boundaries (0,70) due to  small jet inclination (39 deg).  

   \begin{figure}
   \centering
 \includegraphics[width=9cm]{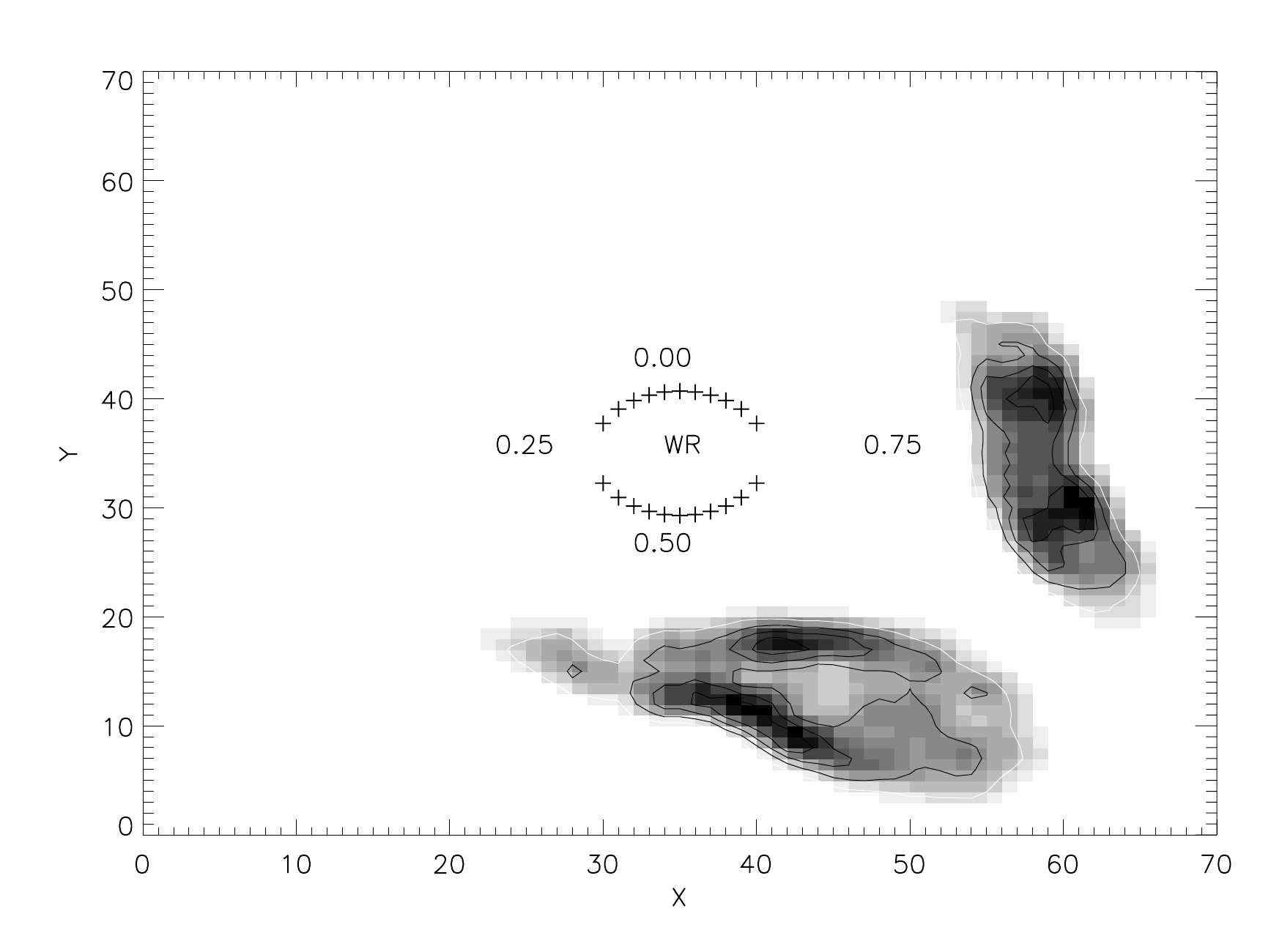}

      \caption{  
     Projections of  the shocked regions in Fig.~2 (assumed to be transformed into clumps)   and advected  by the wind (2000 km/s) during one hour (0.2 $\times$ orbital period).  
      The WR-companion is in the center while the compact  star is moving in a  circular orbit around it (dotted circle). The observation direction is from below (see also Fig.~\ref{crossing}).}
         \label{jetimage2}
   \end{figure}

\section{The ``clumpy trail" and the two absorbers }

We assume that the bow shock triggers the formation of dense clumps or makes the existing clumps in the WR wind denser. The formation of clumps  migh physically similar to star formation  when a shock front impinges upon interstellar clouds  (van Breugel et al.  \cite{vanbreugel}). 
Furthermore, we assume that these clumps participate in the wind outflow with velocity between  1000--3000  km/s.  
Hence, once formed the clumps flow out with  the wind. We stress that we just want to explore the possibility of clumps without  demonstrating
that they really exist.

 We assume that the clump density (like the wind density) decreases with distance r from the WR-star as
 r$^{-\gamma}$ with $\gamma$=2.
The binary separation d = 3.4 R$_{\sun}$  used  in the computations  obviously scales with the wind velocity.
   
 An example of the initial jet bow shock from many computations is shown in Fig.~\ref{jetimage} for a wind velocity 2000 km/s, a cone  with opening angle 20 deg (at orbital phase 0.35), and a tube with radius 0.3d (at orbital phase 0.85). 
 We  further assume that the bow shocks transform 
 wind material  immediately into dense clumps, to be advected by the wind. 
 The  projections of the advected clumps on the orbital plane are shown one hour later in Fig.~\ref{jetimage2}.  
 When coupling similar images over all phases and times we get the outflowing formation we call the ``clumpy trail". Thus the trail is the whole volume of clumps generated by the wind advection.

\begin{figure}
   \centering
 \includegraphics[width=9cm]{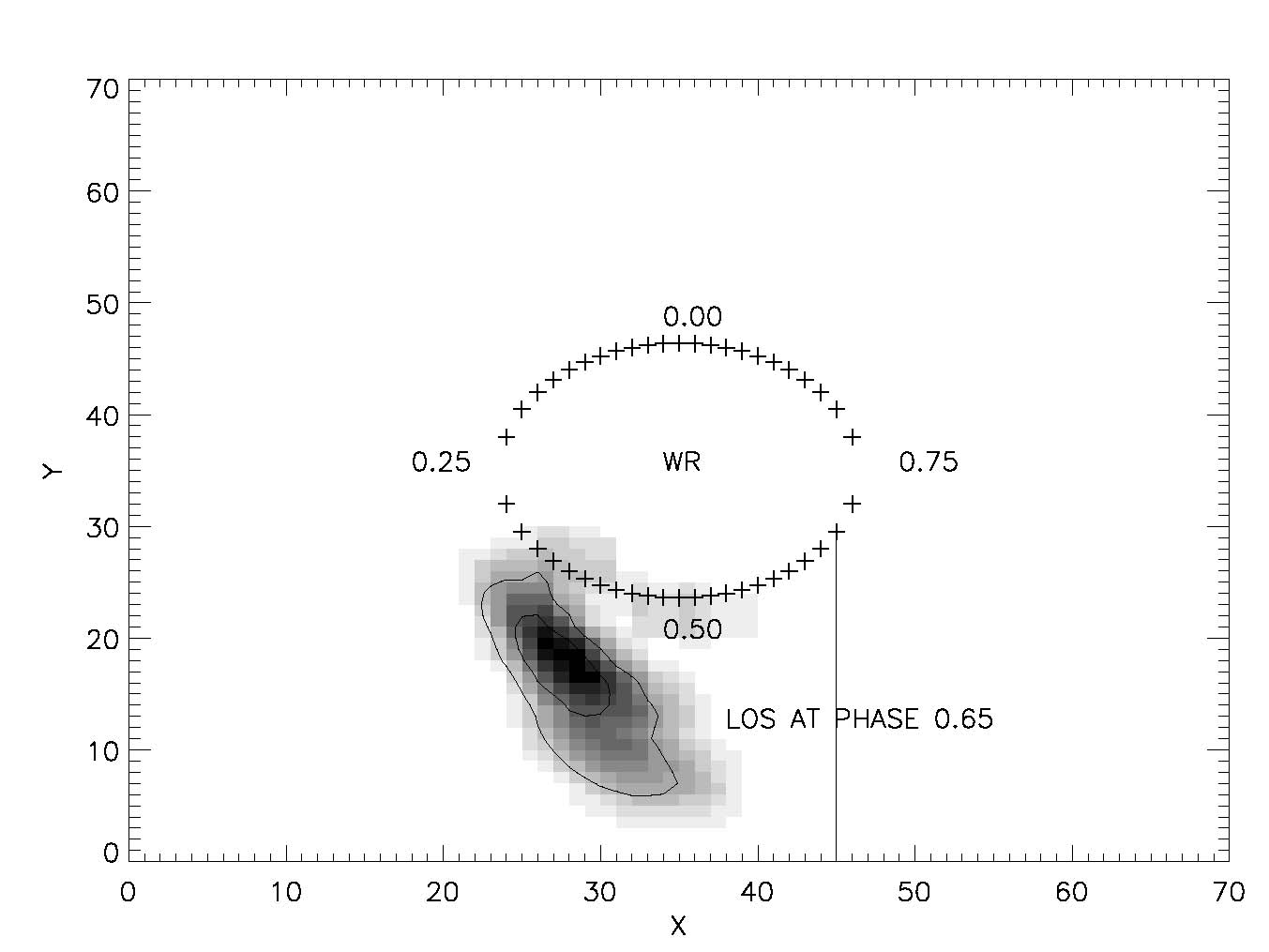}
      \caption{Clump density along the crossing of the line-of-sight  and  clumpy trail (the strip), integrated over all phases and projected on the orbital xy-plane (for the cone in Fig.~\ref{jetimage}). 
      The WR star is in the center, while the compact star is moving in a circular orbit around it  (four orbital phases are shown).  The projection of the  line-of-sight at phase 0.65 is indicated by the vertical line.}
         \label{crossing}
   \end{figure}

The  crossing of  the line-of-sight  (from the compact star) and  the clumpy trail was then computed numerically. For a fixed orbital phase this intersection is a 
 segment ('strip') of the line-of-sight along which the clump density is weighted by $r^{-2}$.  
For the  cone in Fig.~\ref{jetimage} these strips integrated over all phases are shown in Fig.~\ref{crossing} as projected on the orbital plane (note the small inclination of 30 deg used and twice smaller pixel size used in this plot). The crossing region distance r from the compact star is between 0.3--3d (1--10 R$_{\sun}$) at the ionization parameter range log($\xi$)=3.75--2.75 erg cm/sec for density 10$^{12}$ cm$^{-3}$ ($\xi$=L/(Nr$^2$)). 
In this plot we use a smaller pixel size than in Fig.~\ref{jetimage} (0.09d)  to better reveal the details.  

\begin{figure}
   \centering
\includegraphics[width=9cm]{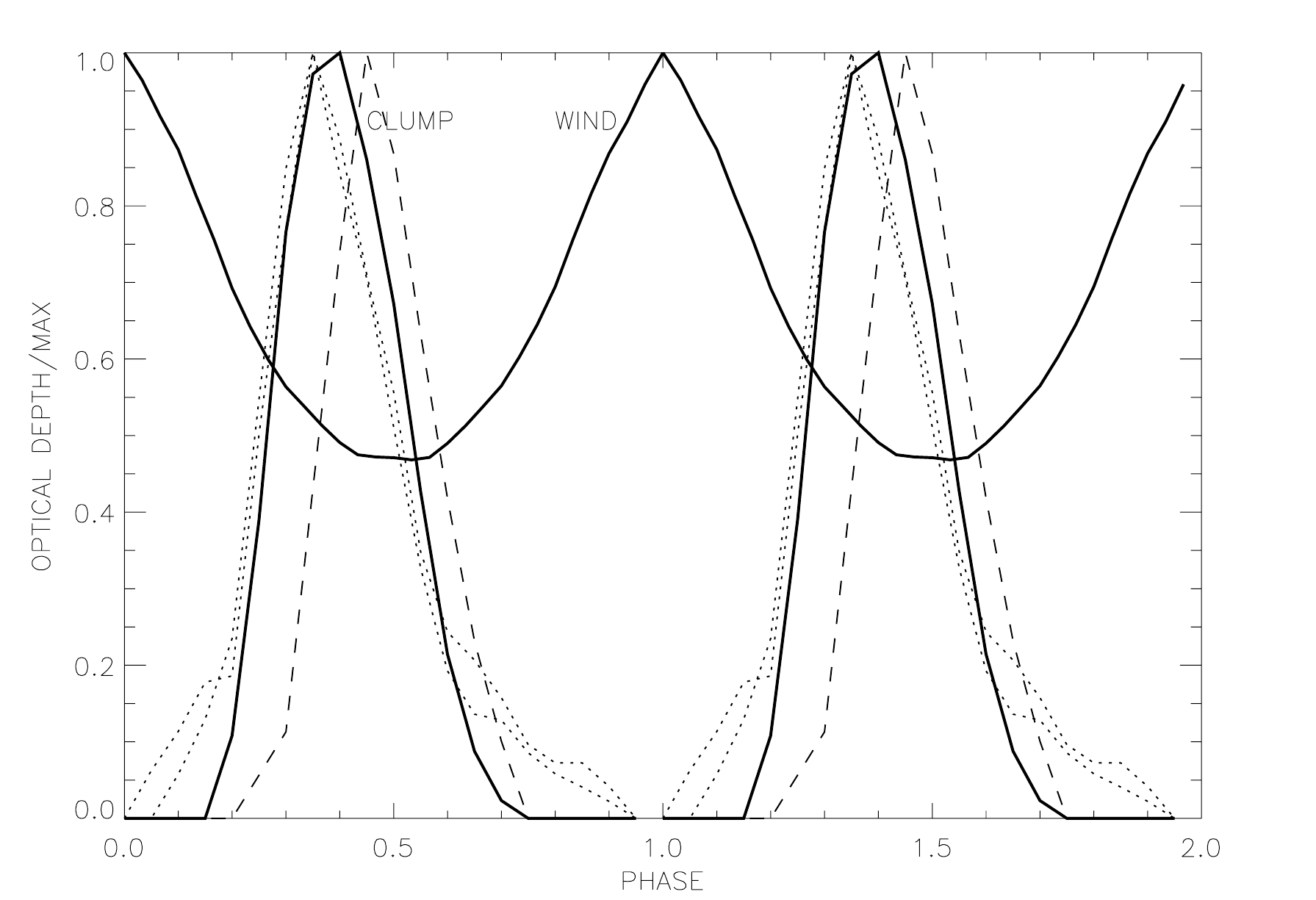}
%
      \caption{ The two absorbers $\tau$(wind) and $\tau$(clump)  used in the modeling in Section 4 (solid lines) where $\tau$(clump)  corresponds in cone geometry to  jet opening angle 20 deg and wind velocity 2000 km/s. The dashed line has opening angle 20 deg and smaller 1000 km/s wind velocity. The two almost coincident dotted lines correspond to a wide  cone  (40 deg, 2000 km/s) and a tube     (see the text in Section 3). } 
         \label{twoabsorbers}
   \end{figure}

Mass columns (assumed proportional to optical depths) along the line-of-sight  for three cone models (20 deg, 1000 km/s), (20 deg, 2000 km/s), (40 deg, 2000 km/s) and a tube with radius 0.3d are shown in Fig.~\ref{twoabsorbers}, scaled with their maximum values along the orbit (dotted lines).  
The average phase at  maxima is around 0.35 while the FWHM is approximately 0.25. 
The  solid line is from the individual cone model in Figs. 2 and 3 (opening angle 20 deg, wind velocity 2000 km/s). 
 Amongst the dozen computed models, the latter produced the best fitting results and  
will be used as the clumpy trail absorber $\tau$(clump) (see Ch 4.)

In the same plot (Fig.~\ref{twoabsorbers}) we include the 
 wind optical depth $\tau$(wind)  (=  mass column
 along the line-of-sight, the  solid line with maximum around phase 0).  This will be used as the first absorber (wind absorber $\tau$(wind)) in Section 4.  
Using (mass column)$^2$ for the optical depth would give a better fit  for the continuum light curve around minimum, and it may  
reflect that $\gamma$ in reality is  larger than 2 (accelerating wind)  or else there is an ionization effect in the wind. 
Using a pure mass column yielded practically the same continuum light curve but with a  partial eclipse profile that was slightly too broad.

Around  $\phi$$_j$ = 39 deg the results are not very sensitive on this jet inclination angle, while changing  $\Theta$$_j$ between 290 deg--340 deg  moves the maximum of $\tau$(clump)  between 0.25--0.40 in phase. 

The key factor in this concept is whether the clumps formed are radiative or not. In our case, the wind (including clumps) is photoionised and  very radiative. Hence, the concept should work (from shocks to dense clumps).

\subsection{Cooling times of shocked clumps}

 Bearing in mind that we have assumed, for the purpose of this paper, 
that a bow shock triggers clump formation, we  nevertheless estimate cooling times for clumps, both in equilibrium with the radiation field as well as heated to 10$^7$ K.  We speculate that hot clumps could exist at short distances from their origin and this is discussed in Section 5.

To estimate  cooling times we use the photoionisation model of Vilhu et al. (\cite{vilhu}),  where the  compact star luminosity  is  L$_{x}$ = 2.46$\times$10$^{38}$ erg/s.  The clump densities probably vary, as do their sizes,  but as an example we use a density of 10$^{13}$ cm$^{-3}$.  Based on  our XSTAR-computations with inclination 30 deg, the wind base density should be around 10$^{12}$ cm$^{-3}$ to guarantee a continuum optical depth around unity  (at 2 keV). This is what the partial eclipse of the continuum light curve requires (see Section 4). A somewhat higher clump density is then a good guess.

\begin{figure}
   \centering
   \includegraphics[width=9cm]{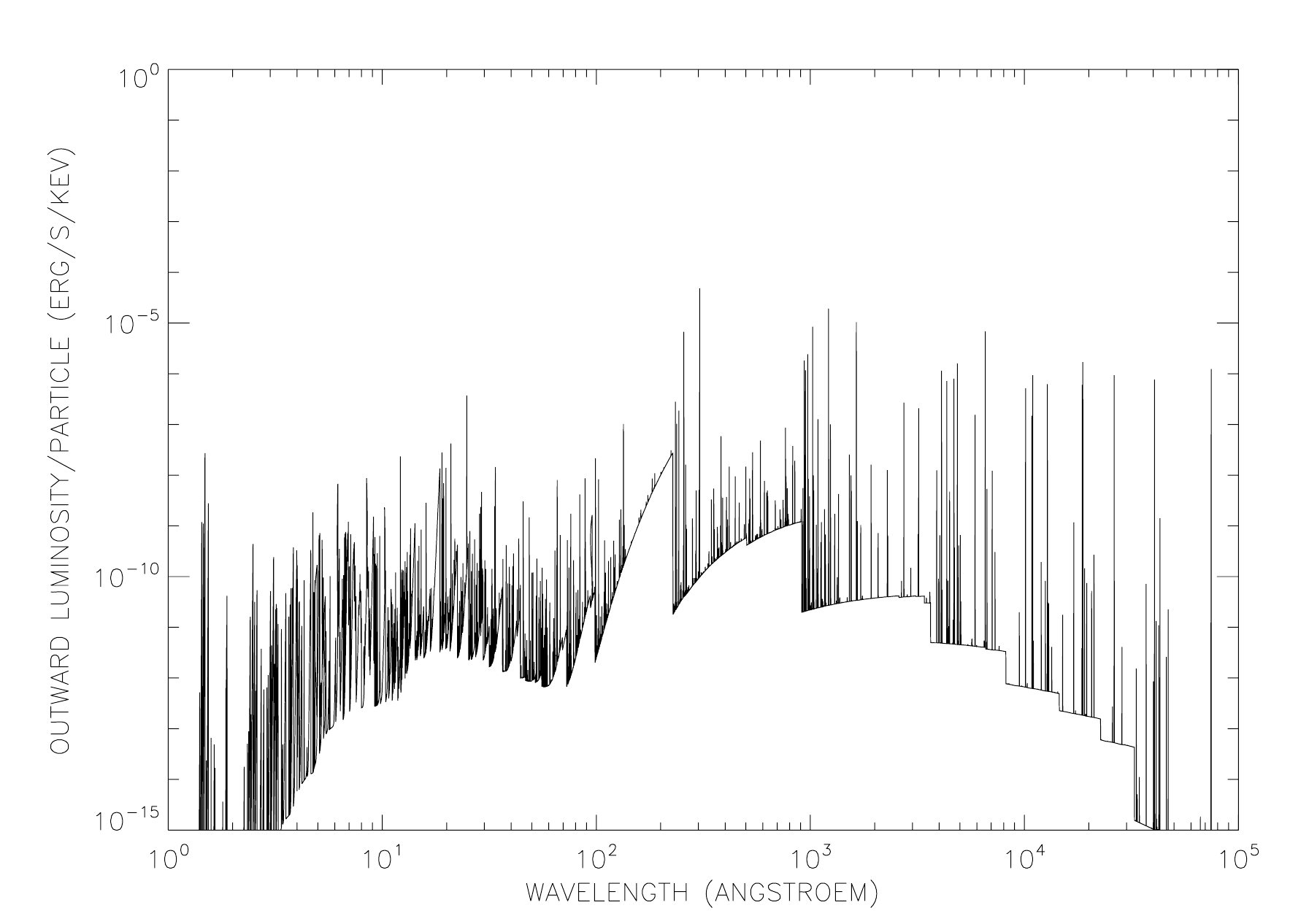}
      \caption{Radiation spectrum of a clump (erg/s/keV per particle)  with density N = 10$^{13}$ cm$^{-3}$ and  in equilibrium temperature (around 10$^{5}$ K) at a distance of 3 R$_{\sun}$  from the compact star (with L$_{x}$ = 2.46$\times$10$^{38}$ erg/s). }
         \label{spectrum}
   \end{figure}

The radiation spectrum of 
 a clump in equilibrium
with density 10$^{13}$ cm$^{-3}$ at a distance of 3 R$_{\sun}$ from the compact  star  (log($\xi$)=3.75)  is shown in Fig.~\ref{spectrum}, as computed with the XSTAR code (Kallman \cite{kallman}).
In this model, the total outward luminosity L$_{out}$  per clump particle equals 4.2$\times$10$^{-9}$ erg/sec, 
while its thermal content per particle in equilibrium with the radiation field (temperature  approximately  10$^{5}$ K) is  1.38$\times$10$^{-11}$ erg.
   
Hence, the cooling time is short (3.3 msec)  but this is 
 counterbalanced
by heating from the radiation field (heating=cooling in equilibrium). 
For 
 a clump 
with N = 10$^{12}$ cm$^{-3}$ the cooling time is 20 msec (with similar equilibrium temperature 10$^{5}$ K).
Cooling times for hot 10$^7$ K clumps (collision dominated) are almost 1000 times longer: 3 sec and  17 sec for densities 10$^{13}$ cm$^{-3}$ and 10$^{12}$ cm$^{-3}$, respectively. 
 This is due to a heat content that is 100 times larger and outward luminosities 10 times smaller of the 10$^7$ K clumps. Cooling times of 10$^8$ K clumps can be several minutes.

Hot 10$^7$ K clumps are not in equilibrium with the surrounding radiation field and try to expand (and cool) after the shock has passed. 
A  crude estimate of the  dynamical time scale can be found as follows.
The force F across the clump surface area S is P$\times$S, where P is the gas pressure inside a clump with mass m. 
From the clump expansion acceleration a=F/m, one can compute  that the dynamical expansion time  is   3.5 sec and 35 sec for clump sizes 10$^8$ cm and 10$^9$ cm, respectively (independent of clump density).  

 Hence, hot 10$^{7}$ K clumps can survive at most a few minutes after their formation, depending on their density and size. Their possible role is discussed in Section 5.

\section{Modeling of the X-ray continuum and emission line light curves}

Here we compare the observed light curves with models using two absorbers: i) the WR wind, and
ii)  an absorber lying in the clumpy trail (see Fig.~\ref{crossing}). 
Both optical depths are plotted in Fig.~\ref{twoabsorbers} and explained in Section 3. 

The observed   gamma-ray ($\it{Fermi}$/LAT) light curve is taken from Abdo et al. (2009) using the pdf measuring perimeter tool for graphics  (Adobe) and modeled with  Dubus et al. (\cite{dubus})  inverse Compton scattering formulas and parameters ($\it{i}$ = 30 deg, $\phi$$_j$=39 deg, $\Theta$$_j$=319 deg, H/d = 2.66). 
The same parameter values were used for the clumpy trail computation. 
Note that Dubus et al. (\cite{dubus}) used phase units where the X-ray minimum occurs at phase 0.25. 
In this paper we use the X-ray phases where the X-ray minimun occurs at phase 0 (the WR star is between the observer 
and the compact star). 
For a jet with $\phi$$_j$=$\it{i}$ and $\Theta$$_j$ = 270 deg, the jet axis and line-of-sight would coincide.

The $\it{RXTE}$/ASM light curve integrated over fifteen years, and scaled inside a specific orbit by the  daily mean, was used as a template for the continuum light curve.  
The ephemeris of Singh et al. (\cite{singh}) was used to compute the orbits. 
The data were limited to moderately high states (daily means $\ge{15}$  counts/sec), to be more consistent 
with the Cygnus X-3 spectral state during the $\it{Chandra}$ HETG observations used.    
The emission line light curves (Si XIV Ly$\alpha$ 6.185 $\AA$,  FeXXVI  Ly$\alpha$  1.780 $\AA$  
 (H-type) and  FeXXV 1.859 $\AA$   (He-type) ) were taken from Vilhu et al. (\cite{vilhu}) and observed during a high state by $\it{Chandra}$/HETG (PI M. McCollough). 
 The number of phase bins used was 30 and 10 for the continuum and lines, respectively.

The modeling of the light curves requires implementation of both the wind and  clumpy trail absorbers explained in previous sections. 
We  assume that the continuum soft X-ray emission is centered at the compact star (disc). 
Most line emission (in particular SiXIV) comes from a broad region in the wind, except  the high excitation FeXXVI Ly$\alpha$ emission that probably originates from the compact star disc (see Vilhu et al.  \cite{vilhu}).
Line absorption in the continuum (at  wavelengths below the line) should then be the common factor for all the lines.   

Let $\tau$(wind) and $\tau$(clump) be the optical depths  along the line-of-sight (from the compact star) in the wind  and clumpy trail, respectively, and scaled with their maximum values along the orbit (as shown in Fig.~\ref{twoabsorbers}).  
As baseline absorbers we adopt those 
 delineated by the heavy solid lines in  Fig.~\ref{twoabsorbers} and  described in Section 3.

The model  flux is defined  with the help of these two absorbers as follows:

\

F = exp[-A$\tau$(wind) - B$\tau$(clump)]  .
                            
\

 A and B  are constants derived from the fitting 
  to observations and  given in Table 1 (using the baseline absorbers). 
The IDL procedure  
 $\it{mpcurvefit.pro}$\footnote{written by Craig Markwardt} was used in the fitting.
The F-statistic and the significance (between 0.0--1.0) of the fit was computed with the IDL-procedure $\it{fv-test.pro}$
 where 1.0 represents the highest significance. 
 Table 1 shows that the fits are significant for both the continuum and the lines. 
If the clumpy trail absorber is not used at all (see Table 2) the fits are not acceptable, in particular for the lines.

\begin{figure}
   \centering
\includegraphics[width=9cm]{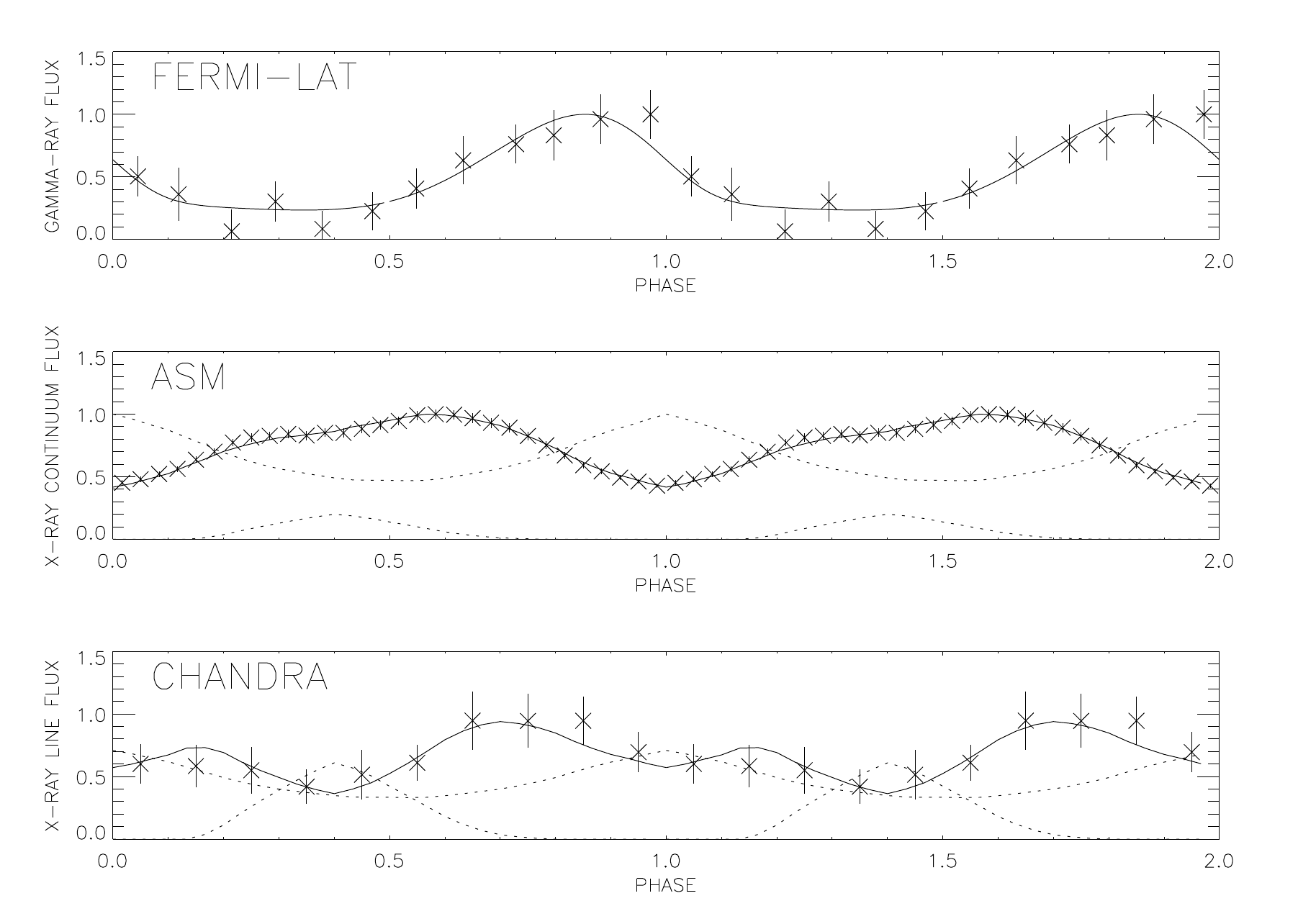}
      \caption{ Observed (crosses) and model (solid)  light curves for  gamma-ray (top, $\it{Fermi}$/LAT),  X-ray continuum (middle, ASM daily mean $\ge{15}$ cps) and   mean of SiXIV, FeXXV and FeXXVI emission lines (bottom, $\it{Chandra}$ HETG). The dotted lines in the two lowest panels show the contributing optical depths of the two absorbers (wind and clumpy trail), scaled with the maximum of continuum wind absorption (the same scale on the left as for the  light curves).}
         \label{lightcurves}
   \end{figure}

Fig.~\ref{lightcurves}  shows the observations overplotted with model fits. 
The mean light curve for the lines is shown,  but note that Table 1 gives fitting parameters for all the lines. 
The contributing optical depths, A$\tau$(wind) and B$\tau$(clump), are also shown in the two lower plots  by dotted lines (scaled with the continuum wind column A$\tau$(wind)).
The scale on the left is the same as for the light curves. 
The gamma-ray  ($\it{Fermi}$ LAT) light curve is shown for comparison in the uppermost panel.

It can be seen that most continuum modulation comes from the wind while the small asymmetry around phase 0.3--0.4  is caused by the clumpy trail (B much smaller than A, see Table 1). 
On the contrary, emission  line absorption is enhanced at the clumpy trail (larger B).
\begin{table}
\begin{minipage}[t]{\columnwidth}
\caption{Fitting parameters and significances for the ASM-continuum (daily means $\ge{15}$ cps) and $\it{Chandra}$ emission line  light curves. DOF is 27 and 7 for the continuum and lines, respectively. MEAN = mean of the lines.} 
\label{fitparam}
\centering
\renewcommand{\footnoterule}{}  
\begin{tabular}{lcccccc}
\hline 
   \\
%
  TYPE  &  A & B     & $\chi$$^{2}$/DOF & F-STAT & SIGNIF \\
\hline
            \\
ASM  &1.82$\pm{0.06}$ & 0.20$\pm{0.03}$   & 0.57 &1.00 &0.99\\
SiXIV &1.97$\pm{0.48}$&1.01$\pm{0.28}$  & 0.62&1.18&0.81\\
FeXXV&0.36$\pm{0.65}$&0.89$\pm{0.40}$   & 0.30&1.25&0.75\\
FeXXVI&1.47$\pm{0.97}$&1.51$\pm{0.58}$   & 0.38&1.39&0.63\\
MEAN&1.29$\pm{0.70}$&1.11$\pm{0.41}$   & 0.30&1.23&0.76\\
\hline 
\end{tabular}
\end{minipage}
\end{table}
\begin{table}
\begin{minipage}[t]{\columnwidth}
\caption{Fitting parameters and significances  for the pure wind absorber ($\tau$(clump)=0 or B=0).} 
\label{fitparampurewind}
\centering
\renewcommand{\footnoterule}{}  
\begin{tabular}{lccccc}
\hline 
   \\
  TYPE  &  A & B     & $\chi$$^{2}$/DOF & F-STAT & SIGNIF \\
\hline
            \\  
ASM  &1.50$\pm{0.04}$ & 0$\pm{0}$   & 2.04 &1.03 &0.93\\
SiXIV &0.55$\pm{0.34}$&0$\pm{0}$  & 2.58&9.1&0.003\\
FeXXV&0.0$\pm{0.0}$&0$\pm{0}$   & 1.76&inf&0.0\\
FeXXVI&0.0$\pm{0.0}$&0$\pm{0}$   & 1.47&inf&0.0\\
MEAN&0.0$\pm{0.0}$&0$\pm{0}$   & 1.40&inf&0.0\\
\hline 
\end{tabular}
\end{minipage}
\end{table}

\section{Discussion}

 The continuum light curve is represented by the ASM average light curve over fifteen years limited to daily means
 $\ge{15}$ counts/sec. 
 So far there is no clear evidence that the form of the light curve changes during this time nor depends on 
 spectral state -- low (hard) or high (soft) -- in a significant way. 
 However, we limited the ASM data to relatively high states to be more consistent with the emission line observations. 

 The emission line  light  curves were observed during a high state and  with  a few days' exposure by $\it{Chandra}$ HETG (Vilhu et al.  \cite{vilhu}, PI M. McCollough). 
 It was assumed that they are representative  also for   the whole set of ASM observations. 
 This is just an assumption and has no observational verification due to the limited amount of  spectral data.  
 The same applies for the $\it{Fermi}$/LAT gamma-ray observations since they are means over one year of observations during flaring periods (Dubus et al.  \cite{dubus}). 

We  assumed that the eventual jet precession has a long time scale (over  tens of years). 
Otherwise the observations could not be compared with the same jet parameters.  
This appears to be the case, since most imaging radio observations between 1991 Jan 15 -- 2001 Sept 15 locate the
 jet position angle in  the  North-South direction (0 deg or 180 deg; Mart\'{i} et al. \cite{marti2}; Mart\'{i} et al.  \cite{marti1}; Miller-Jones et al.  \cite{millerjones}; Schalinski et al. \cite{schalinski}). 
 However,   Tudose et al.  (\cite{tudose}) using  2006 May 01 observations locate knots (which might be interpreted as 
  segments of jets or counter jets) at position angles 55 deg -- 82 deg. 
 Furthermore, Mioduszewski et al. (\cite{mioduszewski})  locate the position angle  during the 1997 Feb 06 observations at 145 deg.
We  also  assumed that the clumpy  trail (in the circumbinary region) is more or less the same during all observations (including big flares or micro-flaring).  
 
The physical nature of the two absorbers requires more work.  
At soft X-rays below 10 keV photoelectric absorption is important and depends both on the distance from the X-ray source and  wind density. 
It also depends  on the element and line transition in question. 
Pure electron scattering (depending on the number of particles along the line of sight) 
contributes in a small way.
To illustrate the situation  we computed 
 three different scenarios with the XSTAR code
(using the X-ray source model explained in Section 3.1) along the line-of-sight from the X-ray source at phase 0.35 (at the strip-segment 0.3d--3d from the  source, see  Fig.~4):

\begin{enumerate}
\item Original wind -- the wind was homogeneous with  density decreasing as r$^{-2}$ with distance r from the WR-star (10$^{12}$ cm$^{-3}$ at the WR-surface

\item Equilibrium clumps -- the wind consisted of equilibrium clumps with density 10$^{13}$ cm$^{-3}$ (in equilibrium with the radiation field)

\item 10$^7$ K clumps -- the clumps were forced to  high  constant temperature 10$^7$ K

\end{enumerate}

\noindent In all  cases the mass columns along the line of sight strip were the same (wind mass conserved). 

Continuum and line optical depths were  computed along the strip using the XSTAR-code. 
Table 3 gives the integrated depths while Fig.~8 gives the optical depths per 10$^{10}$cm versus distance from th X-ray source (in units of the binary separation). 
In the plot the clumps are presented by a mixture (50/50) of equilibrium  
and 10$^7$ K clumps
(solid lines).  This mixture gives a moderate fit to the B-parameter values of Table 1,  provided that the wind base density is twice larger 2 $\times$10$^{12}$ cm$^{-3}$.
The pure wind case is shown by dashed lines. 
The dotted lines give the continuum for clumps (upper line) and wind (lower line) cases. 

Table 3 and Fig.~8 show that the line absorption can indeed be enhaced at the clumpy trail. 
The iron lines appear to require hot collision dominated clumps  (hotter than the equilibrium ones), in particular at short distances from the X-ray source. Replacing the clump density 10$^{13}$ cm$^{-3}$ by 10$^{12}$ cm$^{-3}$ or 10$^{14}$ cm$^{-3}$ does not change much this picture. The short cooling times (see Section 3.1) seem to require some modification to our assumptions: e.g. the jet axis is closer to the line of sight, the jet cone is broader, the bow shock propagates with the wind or clump formation is delayed.
 
\begin{table}
\begin{minipage}[t]{\columnwidth}
\caption{Integrated optical depths along the 'strip' (crossing of  line-of-sight with the clumpy trail) at phase 0.35 (see Fig.~4 and text in Section 5).} 
\label{taus}
\centering
\renewcommand{\footnoterule}{}  
\begin{tabular}{lccc}
\hline 
   \\
  FEATURE  &  ORIG WIND & EQUIL CLUMPS     & 10$^7$ K CLUMPS\\
\hline
            \\  
ASM  & 0.04& 0.13&0.04\\
SiXIV &0.02&0.96& 0.14\\
FeXXV&0.12   & 0.14&1.12\\
FeXXVI&0.13&0.05   & 0.42\\
\hline 
\end{tabular}
\end{minipage}
\end{table}

   \begin{figure}
   \centering
\includegraphics[width=9cm]{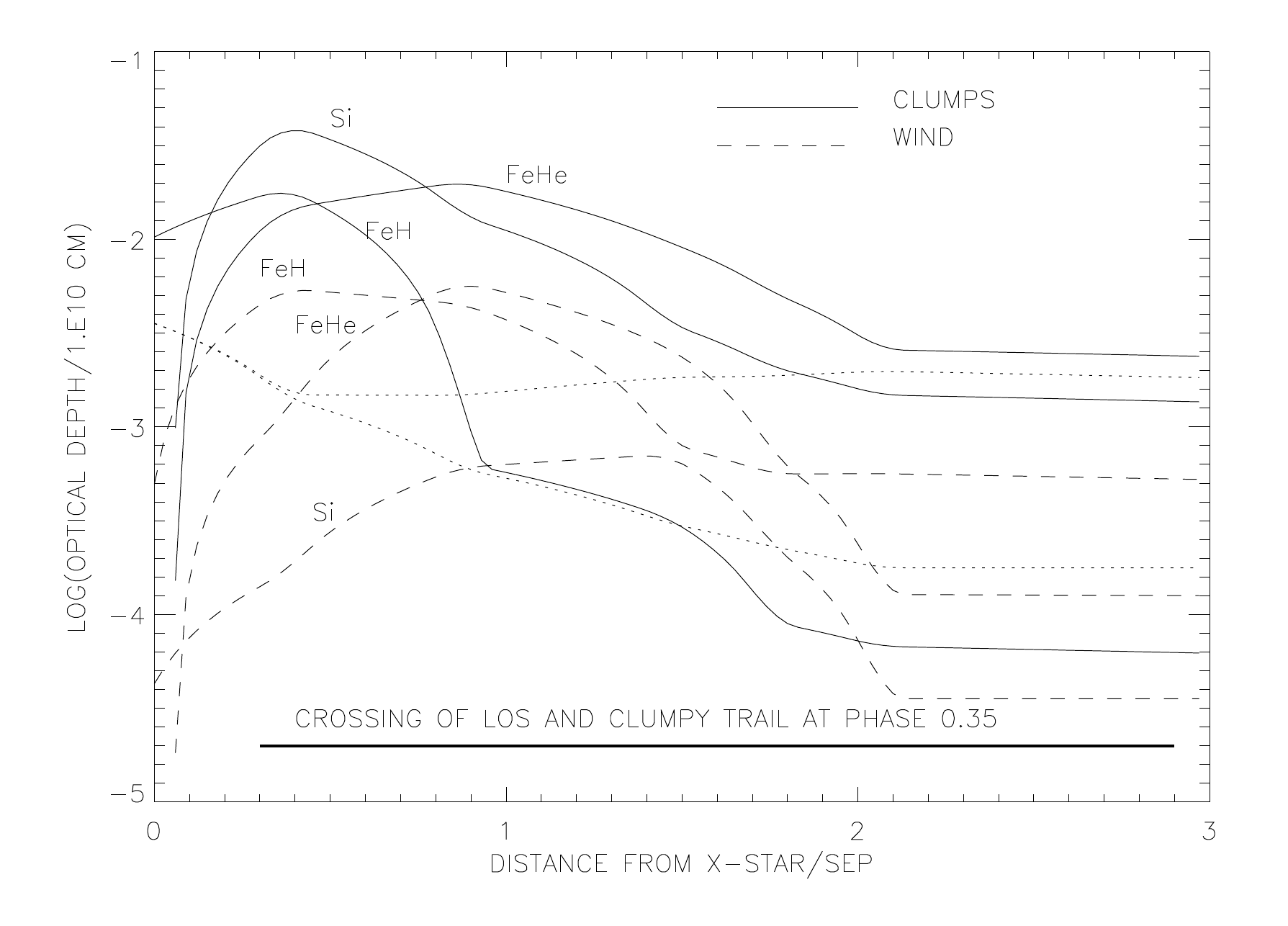}
      \caption{Optical depths vs distance from the X-ray source along the crossing of line-of-sight with the clumpy trail at phase 0.35 (see Fig.4 and  text in Section 5). }
         \label{lastfigure}
   \end{figure}

It is 
noteworthy that the  presence of the clumpy trail absorber occurs around  phase 0.3 (see Fig.~3) which is also where the 9 mHz  QPOs were found by Koljonen et al. (\cite{karri2}). 
Whether these  QPOs arise from  some sort of flickering when the line of sight passes the moving clumps remains to be solved.

 In the present paper  a formal exponential law with two absorbers was used to model the light curves (Section 4). 
 However, it is significant to note that the location of the second  absorber coincides with the clumpy trail region.

\section{Conclusions}

We have demonstrated a possible method to study observable effects of  jets in microquasars when the jet direction and line-of-sight are close to each other (as in Cygnus X-3).
In particular, we showed that the jet in Cygnus X-3  can produce a  ``clumpy trail'' which crosses the line-of-sight
during the binary orbit periodically. 

 Using jet parameters  derived by Dubus et al. (\cite{dubus})  from $\it{Fermi}$/LAT observations, we computed the location of the  clumpy trail  (see Fig.~3). 
 Model light curves were then constructed using the two absorbers in Fig.~5 weighted with constants A and B : i) the WR wind, and ii)  a clumpy trail.  
 These light curves were compared with the observed ones.  
 Good agreements  were achieved for the  soft X-ray continuum ($\it{RXTE}$/ASM)   and  emission lines ($\it{Chandra}$ HETG)  (see Fig.~\ref{lightcurves} and Table 1).
 
We found that the location of the clumpy trail computed from jet parameters matches well with that of the second absorber required (clumpy trail absorber)   which may not be just a coincidence.
Although more work is required to clarify the physical nature of  this clumpy absorber, we suggest that the clumpy trail  consists of a mixture of equilibrium and hot (shock heated) clumps. 

We note that the results are based on an  assumption of constancy of  the jet position angle over the past fifteen years (long precession time) which needs to be confirmed.  
   
\begin{acknowledgements}
We thank the referee for valuable criticism 
 that greatly improved the paper.
We are grateful to Dr. Guillaume Dubus for correspondence and permission to use his plot (Fig.~1). 
We also thank Wiley Publishers for permission to reproduce this figure.
We thank  Dr. Pasi Hakala for useful  comments on the manuscript.   
\end{acknowledgements}


\begin{thebibliography}{}
\bibitem[2009]{dubus}Abdo, A.A. et al. (Fermi-LAT collaboration) 2009, Science, 326, 1512
  \bibitem[2007]{crowther} Crowther, P.A. 2007, 'WR-Stars', Ann.Rev.Astr.Astrophys., 45, 177
\bibitem[2010]{dubus}Dubus, G., Cerutti, B., \& Henri G. 2010, MNRAS, 404, L55
\bibitem[2008]{hamann}Hamann, W-R., Feldmeier, A., \& Oskinova, L.M (editors). 2008, 'Clumping in Hot Star Winds', Proceedings of International Workshop held in Potsdam, Germany, 18-22 June 2007, Potsdam:Univ.-Verl. URN:nbn:de:kolv:517-opus-13981.
 \bibitem[2000]{hanson}Hanson, M.M., Still, M.D., \& Fender, R.P. 2000, ApJ, 541, 308 
 \bibitem[2009]{hjal}Hjalmarsdotter, L., Zdziarski, A.A., Szostek, A., \& Hannikainen, D.C. 2009, MNRAS, 392, 251
  \bibitem[2006]{kallman} Kallman T.R. 2006, 'A Spectral Analysis Tool XSTAR', version   2.1kn6, Goddard Space Flight Center, May25, 2006.
\bibitem[2010]{karri}Koljonen, K.I.I., Hannikainen, D.C., McCollough, M.L., Pooley, G.G., \& Trushkin, S.A. 2010, MNRAS, 406, 307
\bibitem[2011]{karri2}Koljonen, K.I.I., Hannikainen, D.C., McCollough, M.L. 2011, MNRAS, 416, L84
 \bibitem[1989]{langer} Langer, N. 1989, A\&A, 210, 93
 \bibitem[2007]{HMXB} Liu, Q.Z., van Paradijs, J., \& van den Heuvel, E.P.J. 2007, A\&A, 469, 807
\bibitem[2000]{marti2} Mart\'{i}, J., Paredes, J.M., \& Peracaula, M. 2000, ApJ, 545, 939
\bibitem[2001]{marti1} Mart\'{i}, J., Paredes, J.M., \& Peracaula, M. 2001, A\&A, 375, 476
\bibitem[2004]{millerjones}Miller-Jones, J.C.A., Blundell, K.M., Rupen, M.P., et al. 2004, ApJ, 600, 368
\bibitem[2001]{mioduszewski}Mioduszewski, A.J., Rupen, M.P., Hjellming, R.M.,  et al. 2001, ApJ, 553, 766
\bibitem[2008]{perucho}Perucho, M., \& Bosch-Ramon, V. 2008, A\&A, 482, 917
\bibitem[2011]{perucho2}Perucho, M., \& Bosch-Ramon, V. 2012, A\&A, 539, 57
\bibitem[1998]{schalinski}Schalinski, C.J., Johnston, K.J., Witzel, A., et al. 1998, A\&A, 329, 504
\bibitem[2008]{anna}Szostek, A., \& Zdziarski, A.A. 2008, MNRAS, 386, 593
\bibitem[2002]{singh}Singh, N.S., Naik, S., Paul, B., et al. 2002, A\&A, 392, 161
\bibitem[2007]{tudose}Tudose, V., Fender, R.P., Garrett, M.A., et al. 2007, MNRAS, 375, L11
\bibitem[2004]{vanbreugel}van Breugel, W., Fragile, C., Anninos, P., \& Murray S. 2004, 'Jet-induced star formation', in IAU Symposium Series, Vol.217, 472.
\bibitem[1992]{kerk92} van Kerkwijk, M.H.,Charles, P.A., Geballe, T.R. et al. 1992, Nature, 355, 703
\bibitem[1996]{kerk96} van Kerkwijk, M.H., Geballe, T.R., King, D.L. et al. 1996, A\&A, 314, 521
\bibitem[2009]{vilhu}Vilhu, O., Hakala, P., Hannikainen, D.C., McCollough, M., \& Koljonen K. 2009, A\&A, 501, 679
\end{thebibliography}
\end{document}